\documentclass[prd,aps,onecolumn,showpacs,preprintnumbers,amsmath,amssymb,nofootinbib]{revtex4}
\usepackage[dvips]{graphicx}
\usepackage{epsf}
\usepackage{amsmath}
\usepackage{amssymb}
\usepackage{color}

\usepackage{graphicx}% Include figure files
\usepackage{dcolumn}% Align table columns on decimal point
\usepackage{bm}% bold math
\def\be{\begin{equation}}
\def\ee{\end{equation}}
\def\bea{\begin{eqnarray}}
\def\eea{\end{eqnarray}}

\usepackage{color}

\begin{document}

%\date{\today}

\title{Dark Energy Perturbations Revisited}

\author{Mingzhe Li$^{a,d,e}$, Yifu Cai$^{b, f}$, Hong Li$^{b,e}$, Robert Brandenberger$^{c}$ and Xinmin Zhang$^{b,e}$}
\affiliation{a Department of Physics, Nanjing University, Nanjing
210093 P.R. China}
\affiliation{b Institute of High Energy Physics, Chinese Academy
of Sciences, P.O. Box 918-4, Beijing 100049, P.R. China}
\affiliation{c Department of Physics, McGill University,
Montr\'eal, QC, H3A 2T8, Canada}
\affiliation{d Joint Center for Particle, Nuclear Physics and
Cosmology, Nanjing University-Purple Mountain Observatory,
Nanjing 210093, P.R. China}
\affiliation{e Theoretical Physics Center for Science Facilities,
Chinese Academy of Sciences, P.R. China}
\affiliation{f Department of Physics, Arizona State University, Tempe, AZ 85287, USA}


\begin{abstract}
In this paper we study the evolution of cosmological perturbations in
the presence of dynamical dark energy, and revisit the issue of dark
energy perturbations.  For a generally parameterized equation of
state (EoS) such as $ w_D(z) = w_0+w_1\frac{z}{1+z}~,$ (for a single
fluid or a single scalar field ) the dark energy perturbation
diverges when its EoS crosses  the cosmological constant
boundary $w_D=-1$. In this paper we present a method of treating the
dark energy perturbations during the crossing of the $w_D=-1$ surface by
imposing matching conditions which require the induced 3-metric on
the hypersurface of $w_D=-1$ and its extrinsic curvature  to be
continuous. These matching conditions have been used widely in the
literature to study perturbations in various models of early
universe physics, such as Inflation, the Pre-Big-Bang and Ekpyrotic
scenarios, and bouncing cosmologies. In all of these cases
the EoS undergoes a sudden change. Through a detailed
analysis of the matching conditions, we show that $\delta_D$ and
$\theta_D$ are continuous on the matching hypersurface.  This
justifies the method used \cite{Zhao:2005vj,xia1,xia2,xia3}
in the numerical calculation and data
fitting for the determination of cosmological parameters.
We discuss the conditions under which our analysis is
applicable.

\end{abstract}

\maketitle

\section{Introduction}

Since the discovery that the expansion of the universe has recently
been accelerating, a discovery made in particular through
observations of distant Type Ia supernovae (SNIa) in 1998
\cite{Riess:1998cb, Perlmutter:1998np}, a lot of effort has been
made to understand the reason for the acceleration. The most popular
interpretation of the data is to assume that the current universe is
dominated by a new form of matter with negative equation of state
denoted ``dark energy''.  The equation of state (EoS) $w_D$ of the
dark energy, defined as the ratio of its pressure to energy density,
is usually used to classify the different dark energy models. One of
the candidates for dark energy is the cosmological constant whose
EoS $w_D$ is a constant and equals $-1$ at all times. In dynamical
dark energy models extensively discussed in the literature, such as
quintessence\cite{Wetterich:1987fm,Ratra:1987rm,Caldwell:1997ii},
phantom\cite{Caldwell:1999ew},
k-essence\cite{ArmendarizPicon:2000dh}, quintom\cite{Feng:2004ad}
and so on, $w_D$ is generally a function of the redshift
\footnote{For example, see Refs. \cite{Copeland:2006wr,
Albrecht:2006um, Linder:2008pp, Caldwell:2009ix, Silvestri:2009hh,
Cai:2009zp,Leandros} for reviews on dark energy.}. For quintessence
dark energy $-1\leq w_D$, while for phantom $w_D\leq -1$. The
salient feature of quintom dark energy is that its EoS crosses the
phantom boundary set by  $w = -1$.

Given this wealth of theoretical models for dark energy,
it is crucially important to use the accumulated high precision
observational data from SNIa, Cosmic Microwave Background (CMB) and
Large Scale Structure (LSS) surveys to constrain the value of $w_D$ and its
evolution. In this data-driven investigation, one needs to begin by
parameterizing $w_D(z)$ and then fit the parameters introduced to
the data. In recent studies, a popular parametrization of the EoS
is the CPL parametrization\cite{Chevallier:2000qy,
Linder:2002et}:
 $
 w_D(z) = w_0 + w_1\frac{z}{1+z}~,
 $
where $w_0$ and $w_1$ are two free parameters. This model is simple
and has a clear interpretation: $w_0$ is the present value of the
EoS and $w_1$ is its derivative with respect to the scale factor
$a$. In Figure 1 we show the whole parameter space of this model.
Interestingly it can be divided into four regions by the two blue
dotted lines which according to the classifications of the models in
terms of the EoS correspond to quintessence, phantom and quintom A
and quintom B respectively. Both quintom A and quintom B have $w_D$
crossing $-1$. However, they cross the cosmological constant
boundary in a different way. For quintom A, $w_D$ transits from
$w_D>-1$ to $w_D<-1$ as the universe expands, but quintom B does in
a opposite way. The crossing point of the two blue dotted lines
corresponds to the model of the cosmological constant.

The EoS merely reflects the nature of dark energy on the background
of a homogeneous universe. Unless we restrict our attention to the
special case of the cosmological constant, we must take into account
the dark energy perturbations to obtain a consistent and  complete
procedure for data analysis, in particular when fitting cosmological
parameters to the data of CMB and LSS. In fact, it has been shown
that the results obtained by data fitting are quite different
depending on whether one includes or does not include the dark
energy perturbations (for examples, see, \cite{Bean:2003fb,
Zhao:2005vj, Xia:2005ge, Xia:2007km, Komatsu:2008hk,Yeche:2005wn}).

With the energy and momentum density perturbations of dark energy
denoted by $\delta_D$ and $\theta_D$, the perturbation equations are
simple if we assume that dark energy consists of a single perfect
fluid or a single scalar field. In this case, if $w_D$ is restricted
not to cross the line of $-1$, the perturbation equations behave
well. However, such an a-priori restriction on the EoS will yield a
biased result because it excludes most of the parameter space of the
model. Thus, in order not to loose generality, one should do the
global data analysis for the whole parameter space as shown in
Figure 1. However, when $w_D$ crosses the line $-1$, the
perturbations will diverge \cite{Feng:2004ad,Vikman:2004dc,
Hu:2004kh,Caldwell:2005ai}. In fact, in this context of General
Relativity as the theory of space-time, it is impossible to obtain a
background which crosses the ``phantom divide" with only a single
scalar field or a single perfect fluid. This is why the quintom
scenario of dark energy needs to introduce extra degrees of freedom
\cite{Feng:2004ad,Vikman:2004dc, Hu:2004kh, Caldwell:2005ai,
Zhao:2005vj, Li:2005fm, xiaofei}\footnote{For a consistent and
complete proof of the no-go theorem, please see,
\cite{Xia:2007km}.}. And this also implies that the parametrization
of $w_D(z)$ for the background evolution will not be applicable
anymore when considering the perturbations consistently. In turn, it
makes the data analysis more complicated and inconvenient. In order
to keep the maximal generality with the least free parameters for
the parameterized EoS, Ref. \cite{Zhao:2005vj} proposed a method to
deal with the dark energy perturbations during the crossing of the
boundary $w_D = -1$ \footnote{See e.g. \cite{Kunz:2006wc,
Fang:2008sn, Hu:2008zd, Hu:2007pj} for relevant study on the dark
energy fluctuations through the cosmological constant boundary.}. In
this method a small positive parameter $\epsilon$ is introduced
which divides the whole time interval into three regions
corresponding to times when $w_D > -1 + \epsilon$, when $w_D < -1 -
\epsilon$ and when $w_D$ is between $-1 - \epsilon$ and $-1 +
\epsilon$, i.e. the region when it crosses $-1$. In the regions with
$w_D > -1 + \epsilon$ and $w_D < -1 - \epsilon$ the perturbation
equations can be solved easily. In the region when
$-\epsilon<1+w_D<\epsilon$, $\delta_D$ and $\theta_D$ are taken to
be constant so that they are continuous in the whole time range.

In this paper, we will revisit the issue of dark energy
perturbations and will pay particular attention to the treatment of the
perturbations when they cross  the line $w_D=-1$ taking
a different view of point. Our starting point is the general relativistic
matching conditions across space-like hypersurfaces
\cite{Hwang:1991an, Deruelle:1995kd} which are in
turn generalizations of the Israel matching conditions
across time-like hypersurfaces \cite{Israel}. These matching
conditions tell us how the metric and its first derivative are related
on the two sides of a distributional source of matter which leads
to the transition between one solution of General Relativity on
one side of the surface to a different solution on the other side
of the surface.

We consider the space-like hypersurface
$w_D(\eta,~x^i) = -1$ (with $\eta$ denoting
conformal time and $x^i$ the spatial coordinates). The matching
conditions tell us that the induced 3-metric on this hypersurface and
its extrinsic curvature should be continuous across the
matching surface \cite{Hwang:1991an, Deruelle:1995kd}.
These matching conditions have been used widely in
studies of  perturbations in various models of early
universe, such as inflation \cite{Hwang:1991an, Deruelle:1995kd},
pre-big-bang cosmology \cite{Durrer:2002jn}, Ekpyrotic
cosmology \cite{Brandenberger:2001bs, Hwang:2001ga} and
non-singular bouncing cosmologies
\cite{Finelli:2001sr, Cai:2007zv, Cai:2008ed, Cai:2008qw} in which
the EoS undergoes a sudden change. Through the analysis of
matching conditions, we will show in this paper that  - under
certain conditions which will be discussed later - $\delta_D$
and $\theta_D$ are indeed continuous on the matching hypersurface,
thus justifying the method suggested in \cite{Zhao:2005vj}.

The present paper is organized as follows: in Section II we
briefly review the difficulties encountered in single field or
single fluid dark energy models when the EoS crosses $-1$, the
solution to this problem obtained by introducing  quintom model,
and the approach of Ref. \cite{Zhao:2005vj} on how to deal with
quintom dark energy perturbations when fitting to observational
data. In Section III we then study the transfer
of dark energy perturbations across the phantom transition
from the point of view of matching conditions on the
hypersurface of $w_D=-1$.  Using the results
of  this method we then perform a global analysis
of the EoS of dark energy within the framework of the CPL parametrization,
making use of current cosmological observations.
Our numerical results show that the
dark energy perturbations cannot be neglected. In Section IV we
summarize and discuss our results.

\section{General consideration of the dark energy perturbation}

In our analysis, we pick the Conformal Newtonian gauge in which
(in the case of a spatially flat universe), the metric including
linear fluctuations is given by
\begin{eqnarray}
ds^2 \, = \, a(\eta)^2[(1 + 2\Phi)d\eta^2 - (1-2\Psi)\delta_{ij}dx^idx^j]~,
\end{eqnarray}
(we are focusing on scalar perturbations only \footnote{We refer
to Ref. \cite{Mukhanov:1990me} for a comprehensive review of
cosmological perturbation theory.}). The metric perturbations
$\Phi$ and $\Psi$ depend on space and time and describe the small
deviations from a homogeneous Friedmann-Robertson-Walker (FRW)
universe. They are determined by the matter perturbations through
the Einstein equations which take the following form when expanded
to linear order
\begin{eqnarray}\label{einstein}
-k^2\Psi-3{\cal H}(\Psi'+{\cal H}\Phi)&=&4\pi G a^2 \delta\rho~,\nonumber\\
k^2(\Psi'+{\cal H}\Phi)&=&4\pi G a^2(\rho+p) \theta~,\nonumber\\
\Psi''+{\cal H}(2\Psi'+\Phi')+(2{\cal H}'+{\cal H}^2)\Phi+{k^2\over 3}(\Psi-\Phi)&=&
4\pi G a^2 \delta p~,\nonumber\\
k^2(\Psi-\Phi)&=&12\pi Ga^2(\rho+p)\sigma~,
\end{eqnarray}
where ${\cal H}=d\ln a/d\eta$ is the conformal Hubble parameter and
the prime denotes the derivative with respect to  conformal
time. The energy density and pressure perturbations are denoted by
$\delta\rho=\delta T^0_0$ and $\delta p=-1/3 \delta T^i_i$,
respectively. The variable $\theta$ denotes the momentum density
perturbation, which is defined by
\be
(\rho+p)\theta \, = \, ik^i\delta T^0_i \, .
\ee
The shear perturbation $\sigma$ relates to the anisotropic stress through the relation
\be
(\rho+p)\sigma \, = \, \hat{k}_i\hat{k}_j(\delta T^i_j-1/3
\delta^i_j\delta T^l_l) \, ,
\ee
and it vanishes if matter is a perfect fluid or consists of a set of scalar
fields as in the cases considered in this paper. Thus, in the cases considered
here we have $\Psi = \Phi$. Given these considerations, one can obtain
the Poisson equation from the Einstein equations (\ref{einstein}):
\be \label{Poisson}
\frac{k^2}{a^2}\Phi \, = \,
-4\pi G\rho[\delta + \frac{3\mathcal{H}}{k^2}(1+w)\theta]~,
\ee
where $\delta \equiv \delta\rho/\rho$ is the density contrast.

If there are many components of matter, then each species has
its own perturbation variables $\delta_i$, $\theta_i$ and
$\delta p_i$. The total perturbations are given by the
sum over of all species:
\bea
& &\rho\delta \, = \, \sum_i\rho_i\delta_i ~,\nonumber\\
& &(\rho+p)\theta \, = \, \sum_i(\rho_i+p_i)\theta_i ~, \nonumber\\
& &\delta p \, = \, \sum_i\delta p_i~.
\eea
If there are no interactions
beyond gravitational ones among these components, the perturbations for
each species satisfy the individual energy and momentum conservation
laws\cite{Ma:1995ey}
\begin{eqnarray}\label{perturbation0}
 \delta_i' &=&
 -(1+w_i)(\theta_i-3\Phi')-3{\cal{H}}\bigg(\frac{\delta{p_i}}{\rho_i}-w_i\delta_i\bigg)~,\nonumber\\
 \theta_i' &=&
 -{\cal{H}}(1-3w_i)\theta_i-\frac{w_i'}{1+w_i}\theta_i+k^2\bigg(\frac{\delta{p_i}/\rho_i}{1+w_i}+\Phi\bigg)~.
\end{eqnarray}
To solve these two equations we need to know how
$\delta p_i$ depends on $\delta_i$ and $\theta_i$:
\be
\delta p_i \, = \,
c_{si}^2 \delta \rho_i + 3\mathcal{H}(1+w_i)\frac{\rho_i}{k^2}(c_{si}^2-c_{ai}^2)\theta_i~,
\ee
where
\be
c_{ai}^2 \, = \, w_i-w_i'/[3\mathcal{H}(1+w_i)]
\ee
is called adiabatic sound speed in the literature, and $c_{si}$ is the sound
speed defined in the comoving frame of the fluid. For a perfect
fluid $c_{si}=c_{ai}$ and for a canonical scalar field $c_{si}=1$.

For the problem discussed in this paper, we assume that the universe
is filled with only two components, the non-relativistic matter
(including dark matter and baryons) and the dark energy. The matter
perturbation equations are well behaved. But the dark energy
perturbation diverges when its EoS
crosses $-1$. For example, if the dark energy is a single
scalar field, we have the perturbation equations from (\ref{perturbation0})
\begin{eqnarray}\label{perturbation}
 \delta_D' &=&
 -(1+w_D)(\theta_D-3\Phi')-3{\cal{H}}(c_{sD}^2-w_D)\delta_D-3\mathcal{H}\frac{w_D'+3\mathcal{H}(1+w_D)(c_{sD}^2-w_D)}{k^2}\theta_D~,\nonumber\\
 \theta_D' &=&
 -{\cal{H}}(1-3c_{sD}^2)\theta_D+\frac{c_{sD}^2k^2}{1+w_D}\delta_D+k^2\Phi~.
\end{eqnarray}
When $w_D$ crosses $-1$ and  if the sound speed remains
positive, then the second equation of (\ref{perturbation}) becomes
singular and $\theta_D'$ diverges. For a single fluid, when
crossing $-1$ the sound speed
\be
c_{sD}^2 \, = \, c_{aD}^2=w_D-w_D'/[3\mathcal{H}(1+w_D)]
\ee
is divergent and can be arbitrarily negative. This is another
way to see that if the dark energy consists of a single degree
of freedom, its EoS cannot cross the boundary $-1$, otherwise
the perturbation equations become singular and lead to
gravitational instability. Thus, to realize the quintom scenario we
should introduce extra degrees of freedom. The simplest quintom
model is constructed by a combination of a quintessence field and
a phantom field \cite{Feng:2004ad}. We know that the perturbations
of a system including both
quintessence and phantom are stable, so in the quintom model there
is no gravitational instability.

The quintom scenario with multi fluids or multi fields of dark
energy can allow $w_D$ to consistently cross the cosmological
constant boundary. However, it introduces more parameters for data
fitting. Here, we would like to keep the number of free parameters
for the parameterized EoS the same as for a single fluid model. A
technique was developed in Ref. \cite{Zhao:2005vj} to treat the
perturbations during the time interval when the dark energy equation
of state crosses the line $w_D=-1$ with the goal of applying the
procedure to quintom models. Specifically the authors of Ref.
\cite{Zhao:2005vj} introduced a small positive parameter $\epsilon$.
For regions with $w_D > -1 + \epsilon$ and $w_D  <  -1 -\epsilon$,
the dark energy behaves like quintessence and phantom, respectively,
and the perturbations can easily be evolved. In the region of
$-\epsilon < 1+w_D < \epsilon$, Ref. \cite{Zhao:2005vj} assumed
$\delta$ and $\theta$ to be constant, i.e., %%
\be \delta_D' \, = \, 0~\,\,\,\,\,~\theta_D' \, = \, 0
\ee
during this phase. Another way of stating this assumption is
that the values of $\delta_D$ and $\theta_D$ are matched between
the two sides of this interval, that is
\be\label{assumption}
\delta_D|_{+} \, = \, \delta_D|_{-}~\,\,\,\,\,~\theta_D|_{+} \, = \, \theta_D|_{-}~,
\ee
where $\delta_D|_{\pm}$ and $\theta_D|_{\pm}$ represent the
corresponding values when $1+w_D \, = \, \pm \epsilon$. Hence,
$\delta_D$ and $\theta_D$ are continuous throughout. With this
method, in fitting the data one does not need to introduce more
parameters. The numerical calculations have shown that this approach
approximated results obtained using quintom models to a high
precision for values of the parameter $\epsilon$  as small as
$10^{-5}$ \cite{Zhao:2005vj,xia1,xia2,xia3}.

\section{Dark energy perturbation with parameterized EoS and matching conditions}

The method to deal with dark energy perturbation with its EoS
across $-1$ proposed in \cite{Zhao:2005vj} assumes that in the
neighborhood of the crossing point the energy and momentum density
perturbations $\delta_D$ and $\theta_D$ are frozen. This
guarantees the continuity of $\delta_D$ and $\theta_D$. In this
section we will investigate this treatment from a different point of view.

Consider a space-like hypersurface $\Sigma$
%%~\,\,\, w_D(\eta,~x^i) \, = \, -1
which divides space-time into the two regions $w_D \geq
w_+=-1+\epsilon$ and $w_D \leq w_-=-1-\epsilon$. The surface
represents the region $-1 + \epsilon > w_D > -1 - \epsilon$, the
region in which the evolution of dark energy fluctuations is not
under control for EoS with a single component. To this surface we
apply the matching conditions of \cite{Hwang:1991an,
Deruelle:1995kd} which state that the induced 3-metric on this
hypersurface and its extrinsic curvature are continuous.

These matching conditions can be applied to reheating
in inflationary cosmology: instead of solving the
equations of motion in a specific model which describes
the transition from the inflationary phase to the radiation
phase after reheating, we cut out a time interval
$t_R - \epsilon < t < t_R + \epsilon$ about the
reheating time $t_R$ and apply the matching conditions
to connect the fluctuations on either side of this
interval. Similarly, these matching conditions have
been applied to pre-big-bang and Ekpyrotic cosmology
to cut out a time interval about the time when the
background is singular and then connect the fluctuations
on either side of the matching surface.

%%RB New paragraph
As pointed out in \cite{Durrer:2002jn}, this matching procedure is
not well justified if the background does not obey
the matching conditions. Therefore, in recent
studies of non-singular bouncing cosmologies
\cite{Cai:2007qw}
one introduces a bouncing phase valid
around the bounce point and matches both at
the boundary between the initial contracting
phase and the onset of the bouncing phase,
and then once again between the end of the
bouncing phase and the final expanding
phase. In this case, the fluctuations can also
be evolved numerically and one can
verify that the approximate analytical
description of the evolution of fluctuations using
matching conditions gives accurate results for the
evolution of cosmological perturbation. Note that
in this case the equation of state of the background also
has $w$ crossing $-1$ at the bounce point \cite{Cai:2008qb}.

%%RB New paragraph
However, in our present investigation the matching
prescription for fluctuations is justified since the
background satisfies the corresponding conditions. Thus,
it is sufficient to use a single matching surface, like in
the case of inflationary reheating.

In a homogeneous universe, the matching hypersurface coincides
with that of fixed conformal time $\eta$. In the presence of small
amplitude inhomogeneities the EoS can be decomposed into a
homogeneous part and a small perturbation:
\be
w_D \, = \, w_D(\eta) + \delta w_D(\eta,~x^i) \, .
\ee
To obtain the matching conditions on this
hypersurface, it is better for us to consider the general form of
the perturbed metric
\be
ds^2\, = \, a^2(\eta)\{(1+2A)d\eta^2-2B_{,i}dx^id\eta-[(1-2\psi)\delta_{ij}+2E_{,ij}]dx^idx^j\}~,
\ee
where commas denote derivatives with respect to spatial
coordinates. Only two of the four variables $A,~B,~\psi,~E$ are
physical. Under the coordinate transformation
\bea
\eta \, & \rightarrow & \, \tilde{\eta} = \eta+\xi^0  \,\,\,\, {\rm and} \nonumber \\
x^i \, & \rightarrow & \,
\tilde{x}^i=x^i+\xi^{,i} \, ,
\eea
these metric perturbations transform as
\bea\label{gauge}
& &A\rightarrow \tilde{A} \, = \, A-\mathcal{H}\xi^0-\xi^{0'}~,~~~
B \, \rightarrow \, \tilde{B} = B + \xi^0 - \xi'~,\nonumber\\
& &\psi \, \rightarrow \,
\tilde{\psi} = \psi + \mathcal{H}\xi^0~,~~~~~~~~~
E \, \rightarrow \, \tilde{E} = E - \xi~,
\eea
and the perturbation of the EoS transforms as
\be
\delta w_D \, \rightarrow \,
\widetilde{\delta w}_D = \delta w_D - w_D'\xi^0~.
\ee

We will use the temporal gauge to obtain the
matching conditions. In this gauge, the matching hypersurface
$\Sigma$ coincides with $\tilde{\eta} = {\rm const.}$ and the
equation of this hypersurface
\be
\widetilde{w}_D(\tilde{\eta},~\tilde{x}^i) \, = \, {\rm const.}
\ee
implies
\be
\widetilde{\delta w}_D \, = \, 0 \, .
\ee
Hence the time shift is
\be \label{gauge2}
\xi^0 \, = \, \frac{\delta w_D}{w_D'}~,
\ee
but $\xi$ remains arbitrary. The induced 3-metric of this
hypersurface and its extrinsic curvature are expressed as
\bea
q_{ij} \, &=& \, a^2[(1-2\tilde{\psi})\delta_{ij}+2\tilde{E}_{,ij}]~, \\
K_{ij} \, &=& \, \frac{q_{ij}}{a}(\mathcal{H}-\mathcal{H}\tilde{A}-\tilde{\psi}')+\frac{1}{a}(\tilde{E}'-\tilde{B})_{,ij}~,
\eea
respectively.

The matching conditions tell us that the induced metric and the
extrinsic curvature should be continuous across the surface, i.e.
that $[q_{ij}]_{\pm} = 0$ and $[K_{ij}]_{\pm} = 0$. For the
background, this requires that the scale factor $a$ and the
expansion rate $\mathcal{H}$ are continues. And for the
perturbations, one obtains %%
\be
[\tilde{\psi}]_{\pm} \,  =  \, [\tilde{E}]_{\pm} \, = 0 \,
\ee
and
\be
[{\mathcal{H}} \tilde{A}  +  {\tilde{\psi}}^{'} ]_{\pm}
\, = \, [ {\tilde{E}}^{'}  -  {\tilde{B}} ]_{\pm}
= 0 ~,
\ee
where the notation
\be
[\tilde{\psi}]_{\pm} \, \equiv \,
\tilde{\psi}_+-\tilde{\psi}_- {\rm etc.}
\ee
(the subscripts $+$ and $-$ indicating the values of the
quantity on the two sides of the boundary) has been used. Making use of the gauge
transformations (\ref{gauge}) and (\ref{gauge2}), we obtain the
matching conditions for the perturbations in an arbitrary gauge,
\bea\label{match}
& &[\psi+\mathcal{H}\frac{\delta w_D}{w_D'}]_{\pm}=0~ ,\nonumber\\
& &[E-\xi]_{\pm}=0~ ,\nonumber\\
& &[\mathcal{H}A+\psi'+(\mathcal{H}'-\mathcal{H}^2)\frac{\delta w_D}{w_D'}]_{\pm}=0~ ,\nonumber\\
& &[E'-B+\frac{\delta w_D}{w_D'}]_{\pm}=0~.
\eea

Specifically, in the Conformal
Newtonian gauge used in this paper ($B = E = 0$ and
$\Phi=A$, $\Psi=\psi$) these conditions become
\bea\label{match2}
& &[\Psi]_{\pm} = 0 ~, \nonumber\\
& &[\frac{\delta w_D}{w_D'}]_{\pm} = 0 ~, \nonumber\\
& &[\mathcal{H}\Phi+\Psi'+(\mathcal{H}'-\mathcal{H}^2)\frac{\delta
w_D}{w_D'}]_{\pm} = 0~.
\eea
When $\Phi=\Psi$, i.e. in the absence of shear perturbations, and
dividing the matter contributions into that of dark energy and that of
regular cold matter,  the Poisson equation (\ref{Poisson}) becomes
\be
\frac{k^2}{a^2}\Phi \, = \, -4\pi
G\{\rho_D[\delta_D+(1+w_D)\frac{\mathcal{H}}{k^2}\theta_D]+\rho_m[\delta_m+(1+w_m)\frac{\mathcal{H}}{k^2}\theta_m]\}~,
\ee
where the subscript $m$ denotes matter.

The first matching
condition in (\ref{match2}) means that the combination
$\delta_D+(1+w_D)\frac{\mathcal{H}}{k^2}\theta_D$ should be also
continuous. Because the matching hypersurface is characterized by
$w_D=-1$, one gets the following matching condition for the energy density
perturbation of dark energy
\be \label{match3}
[\delta_D]_{\pm} \, = \, 0~.
\ee

Now we turn to the physical meaning of the second condition in
(\ref{match2}). After simple calculations one gets
\be \frac{\delta
w_D}{w_D'} \, = \, \frac{1}{w_D'}(\frac{\delta p_D}{\rho_D}-w_D\delta_D) \,
= \, \frac{c_{sD}^2-w_D}{w_D'}[\delta_D+\frac{3\mathcal{H}(1+w_D)}{k^2}\theta_D]+\frac{\theta_D}{k^2}~,
\ee
and at the matching hypersurface this becomes
\be
\frac{\delta w_D}{w_D'} \, = \, \frac{c_{sD}^2+1}{w_D'}\delta_D+\frac{1}{k^2}\theta_D~.
\ee
Both $\delta_D$ and $w_D'$ are continuous, and $w_D'$ must be non-zero
in order to obtain crossing. Thus, the matching condition
$[\delta w_D/w_D']_{\pm}=0$ implies that the momentum density perturbation of dark
energy is also continuous, i.e.
\be \label{match4}
[\theta_D]_{\pm} \, = \, 0~.
\ee
Eqs. (\ref{match3}) and (\ref{match4}) coincide with the assumptions
(\ref{assumption}) used in Ref. \cite{Zhao:2005vj}. Another way to see
that (\ref{match3}) and (\ref{match4}) are valid we see that if these
matching conditions are satisfied, then all of the matching conditions
(\ref{match2}) are satisfied.
%%RB Extra sentence added above.

\begin{figure}[htbp]
\begin{center}
\includegraphics[scale=0.5]{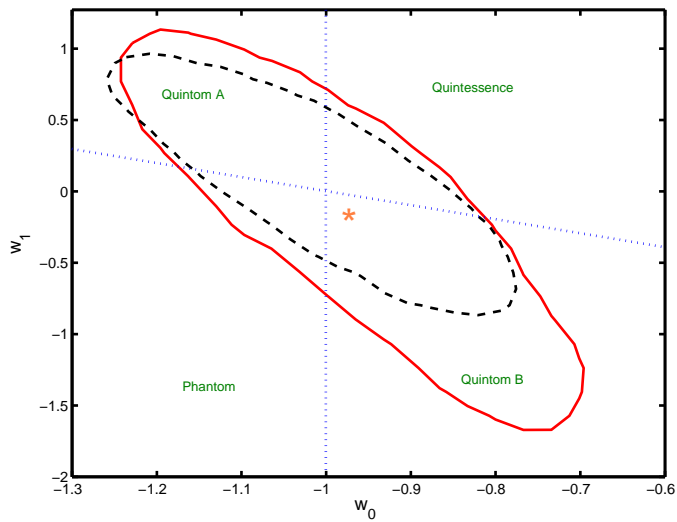}
%\centerline{\epsfxsize=\figsize\epsffile{agelmdaSN.eps}}
\vskip-1.3cm \vspace{10mm}\caption{Two dimensional constraints on
$(w_0, w_1)$ from  current observations of CMB $+$ SNIa $+$ BAO. The
red solid and black dash lines represent the 2 $\sigma$ limits for
the two cases with and without dark energy perturbations,
respectively. The star represents the best fit value.
\label{figw0wa}}
\end{center}
\end{figure}

Now, with the method discussed in this paper we can perform a
numerical calculation to see how large the contribution of the dark
energy perturbation can be. We modified and extended the CosmoMC
code by implementing the dark energy perturbations discussed in this
paper and take $\varepsilon = 10^{-5}$, then fit the parameters of
the dark energy EoS ($w_0, w_1$) to the current data from  CMB
observations including the 7-year WMAP temperature and polarization
power spectra \cite{Komatsu:2010fb}, and small-scale CMB
measurements from BOOMERanG \cite{BOOMERanG}, CBI \cite{CBI}, VSA
\cite{VSA} and ACBAR \cite{ACBAR}, from the Union2 SNIa data set
\cite{union2}, and from BAO\cite{BAO}. In order to show the
importance of the dark energy perturbation we have done the
calculations separately for the two cases including and switching
off the dark energy perturbations. In Figure \ref{figw0wa} we plot
our numerical results. One can see the obvious difference between
the two cases given by the red solid line and the black dashed line.
This is because the late time ISW effect differs significantly when
dark energy perturbations are considered, and the ISW effects plays
an important role on large angular scales for the CMB and the matter
power spectra \cite{Li:2008cj}.

\section{Summary and Discussion}

In this paper we have revisited the dynamics of cosmological
perturbations of dark energy and paid particular attention to the
case when the EoS crosses the cosmological constant boundary.
Single field or single fluid models, or scenarios based on
a parameterized EoS of dark energy with a single component
cannot cross $w_D = -1$ because the perturbations
are singular and unstable at this point. The quintom model is able to
cross this boundary naturally, however it requires  more degrees
of freedom, and lessons learned when studying the transfer
of fluctuations through non-singular bounces makes us expect
that, on scales smaller than the time duration of the transition
phase, the final fluctuations will depend on the details of the
model. This makes it hard to obtain a simple data fitting
prescription. In particular, the more parameters are
introduced, the more computing time is required for the numerical
calculations.  To obtain a simple way of analyzing data
and assessing the observational evidence for or against the
equation of state of dark energy crossing the cosmological
constant divide it is thus very useful to have a prescription
which does not introduce new parameters.

In this paper we have presented a new approach to studying  dark
energy perturbations in the time interval when [ $w_{+}= -1 +
\epsilon, ~ w_{-}= -1 -\epsilon$ ] , i.e.  during the crossing of
the boundary $w_D =-1$. We have proposed to apply the general
relativistic matching conditions of \cite{Hwang:1991an,
Deruelle:1995kd}. These conditions imply that the dark energy
perturbations match continuously on the two sides of the surface
$w_D= -1$.

%%RB New paragraph
Let us mention some caveats to our analysis: Our method is
applicable in the form presented here only if on either side of the
matching surface all except for one fluid are negligible. Since at
the crossing region this assumption will fail, this criterium
implies that $\epsilon$ cannot be too small. Secondly, since the
dark energy fluctuations diverge when the equation of state crosses
the cosmological constant line, then, in order to stay within the
realm of applicability of linear cosmological perturbation theory,
we have a second reason why $\epsilon$ cannot be taken to be too
small. On the other hand, for length scales smaller than $\epsilon
t$, where $t$ is the time when the EoS of dark energy crosses the
cosmological constant line, the way in which the fluctuations pass
through the transition region may depend on the specific quintom
models. This argument prefers a small value for $\epsilon$.

However, let us consider models where our assumptions are satisfied
and where $\epsilon$ is sufficiently small. Then, our results
coincide with those of Ref. \cite{Zhao:2005vj}, and the arguments in
this paper justify the method used in the numerical calculations of
\cite{xia1, xia2, xia3}. Since linear perturbation theory will break
down as $\epsilon \rightarrow 0$,  the small positive parameter
$\epsilon$ would not be taken to be too small.  In Ref.
\cite{Zhao:2005vj} it has been checked that with $\epsilon \sim
10^{-5}$, linear perturbation theory will be valid and at the same
time the approximation of taking $\delta_D$ and $\theta_D$ to be
constant in the interval [$w_{+}= -1 + \epsilon,  ~ w_{-}= -1
-\epsilon$ ] will yield results in agreement with those obtained by
using actual quintom perturbations.

Finally, with the method outlined in this paper we have performed a
numerical  determination of cosmological parameters. Our numerical
results show explicitly the significance of the dark energy
perturbations.

\section{acknowledgments}

We thank Jun-Qing Xia and Gong-Bo Zhao for helpful comments and
discussions. M.L. is supported by the Specialized Research
Fund for the Doctoral Program of Higher Education (SRFDP) under
Grant No. 20090091120054. Y.C., H.L. and X.Z. are supported in part
by the National Natural Science Foundation of China under Grants Nos.
10975142, 10821063 and 10803001 and by the 973 program Nos.
1J2007CB81540002 and 2010CB833000 and by the Youth Foundation of the
Institute of High Energy Physics under Grant No. H95461N. R.B. is
supported in part by funds from NSERC and from the CRC program of
Canada. He also wishes to acknowledge the warm hospitality of the
cosmology group at the Institute of High Energy Physics during a visit
when this project was started.

\end{document}